# Fabrication of polyhedral particles from spherical colloids and their self-assembly into rotator phases

Hanumantha Rao Vutukuri *, Arnout Imhof & Alfons van Blaaderen *

**Abstract:** *Particle shape is a critical parameter that plays an important role in self-assembly, for example, in designing targeted complex structures with desired properties. In the last decades an unprecedented range of monodisperse nanoparticle systems with control over the shape of the particles have become available. In contrast, the choice of micron-sized colloidal building blocks of particles with flat facets, i.e., particles with polygonal shapes, is significantly more limited. This can be attributed to the fact that, contrary to nanoparticles, the larger colloids are significantly harder to synthesize as single crystals. Herein, we demonstrate that the simplest building block, such as the micron-sized polymeric spherical colloidal particle, is already enough to fabricate particles with regularly placed flat facets, including completely polygonal shapes with sharp edges. As an illustration that the yields are high enough for further self-assembly studies we demonstrate the formation of 3D rotator phases of fluorescently labelled, micron-sized and charged rhombic dodecahedron particles. Our method for fabricating polyhedral particles opens a new avenue for designing new materials.*

Building blocks consisting of non-spherical and anisotropic particles offer a vast variety of structures with different symmetries, packing densities, and directionalities as compared to structures that are built from isotropic spherical particles.[1] In the field of nanoparticles the amount of new particle model systems with a well-defined complex shape has increased exponentially.[1a, c] The methods reported include photochemical, thermal, electrochemical, and template-directed methods but almost exclusively deal with nano-crystalline particles.[1a, c] Now that it is often possible to predict what structure is needed for certain material properties the big challenge is to experimentally realize those particle interactions and protocols which lead to the desired structures. Self-assembly of complex shapes with complex interactions provides one possible route to realize the desired structures. Recently, a growing demand for micron-sized colloidal particles with complex shapes is largely driven by applications in photonics, as well as in advanced functional materials

[∗]  Dr. H.R. Vutukuri, Dr. A. Imhof, Prof. dr. A. van Blaaderen
Soft Condensed Matter, Debye Institute for Nanomaterials Science, Utrecht University, Princetonplein 1, 3584 CC, Utrecht ,The Netherlands
E-mail: H.R.Vutukuri@uu.nl, A.vanBlaaderen@uu.nl
Homepage : http://www.colloid.nl/

Supporting information for this article is given via a link at the end of the document.

design.[1f,g] Several synthesis routes have been reported to fabricate particles with non-spherical geometries. These approaches make use of self-assembly,[2] photolithography,[3] microfluidics,[4] nonwetting template molding,[5] stretching polymer-embedded particle films,[6] thermal sintering of spherical particles at an oil-water interface,[7] *template method,*[1g, 8] *metal-organic frame works,*[9] and seeded emulsion polymerization.[10] Collectively, these methods have produced particles of several distinct shapes, including metal-based polyhedral particles. Moreover, flat facets on the particle surface play an important role in self-assembly. For instance, particles tend to align along the flat interfaces at higher densities,[11] especially for hard particle interactions. With a depletant added this particle geometry enhances directional attractive interaction between two flat interfaces.[3,12] Recent experimental work also indicates that similar arguments hold for van der Waals, electric double-layer and electric field induced interactions.[13] Several recent computational and theoretical studies have predicted that polyhedral colloidal particles show a rich phase behavior including quasicrystals and rotator phases that still allow full particle rotations on a 3D ordered lattice of the polygons,[1e, h] but experimental realizations of those phases with the micron-sized colloidal particles scarcely exist. We illustrate in this paper the methodology that we developed for studying these systems quantitatively on the single particle level.

In the past few years, we developed a thermal sintering method for creating permanently bonded 1D colloidal analogues of polymer bead chains,[2a,b] 2D sheets,[14] and 3D structures[15] with sterically stabilized and 'un-locked particles' (i.e., the stabilizer molecules are not covalently linked to the particles) of polymethyl-methacrylate (PMMA).[16] For the particle deformation procedure presented here to yield individual particles, we found it to be essential that the steric comb-graft type stabilizer molecules be covalently linked to polymer chains in the core of the particles (using a so-called locking step).[16] Sintering phenomena which lead to particle deformation are known to occur in the intermediate stages of latex film formation,[17] which is used in several industrial processes associated with, e.g., paints, paper coatings, textiles, and carpets. The deformation of the particles takes place once they are in close contact with their neighbors either in a dry state (dry sintering) and/or in a wet state (wet sintering).[17] Generally in those stages also some exchange of material from one particle to its neighbors by polymer chain diffusion occurs, but mostly in the final stages of film formation. This effectively prohibits the detachment of the deformed particles and the preservation of any smooth surfaces that have formed.[17-18] Here, we demonstrate that under certain conditions the sintering process can be carried out without any exchange of material among adjacent particles using sterically stabilized and locked particles,[16a] while at the same time effecting the deformation of particles by contact with their direct neighbors. Our method consists of three simple steps (Figure 1): i) the self-assembly of sterically stabilized and locked polymeric spherical particles into a 3D crystalline structure, ii) the deformation of the spherical particles by thermal annealing in such a way that mass transport between the particles does not take place, and iii) the redispersion of the resulting deformed particles from the 3D superstructures into individual particles by sonication. Moreover, we demonstrate the self-assembly of rhombic dodecahedron particles into rotator phases at different salt concentrations.

Several methods are available for growing colloidal crystals. We chose a simple method, crystallization induced by sedimentation, for our nearly refractive-index matched, hard-sphere-like[15,19] colloidal suspension of rhodamine isothiocyanate (RITC) labelled PMMA particles in cyclohexyl bromide (CHB). The suspension, consisting of 10-20% by volume of particles was transferred to a capillary cell with a 0.2 mm X 2.0 mm cross section and of a desired length (~ 10 cm) oriented in an upright position. After 1-2 weeks, a large random hexagonal closed packed crystal was observed. Although face-centered cubic (FCC) is the thermodynamically most stable phase, the free energy difference with respect to the metastable hexagonal-closed-packed (HCP) is very small (~$10^{-4}$ $k_B T$ / particle) at the melting transition. Therefore, a random hexagonal closed packed structure has been observed experimentally more often than pure FCC. However, it is known that particles with a slightly soft potential or sedimentation of hard spheres on a template can be used to grow pure FCC colloidal crystals if needed.[20]

Next, the crystals were heat treated by immersing the capillary in a hot water bath at 75 °C, which is well below the glass transition temperature ($T_g$ = 140-145 °C) of bulk PMMA.[15] In wet sintering, the interfacial tension between the particles and the solvent is the dominant driving force. Moreover, this liquid modifies not only the surface tension, but also the glass transition temperature of the PMMA as it slightly swells the particles.[15] In addition, it is known that even for homogeneous polymer particles the glass transition temperature for the outer layer of the polymer can differ significantly (tens of degrees) from that of the interior.[18a] As a consequence, the touching spheres underwent a plastic deformation in the presence of the solvent. This sintering process has been discussed in literature using a range of models as arising from either a viscous flow process driven by surface tension effects or an elastic Hertzian deformation of elastic spheres under tension.[17-18] Almost certainly the reality is more complex and is likely of a visco-elastic intermediate form where the details depend mostly on particle and stabilizer properties.[17-18]

During the initial stages of sintering, approximately 2-3 minutes after the heating was started, the particles were found to form flat facets at the contact points due to the surface forces between the particles and as mediated by the solvent. Samples were subsequently cooled down to room temperature and dried after opening of the capillary for about 2-3 days at room temperature. It can clearly be seen that the crystal building blocks then consisted of non-spherical particles with small patches as shown in Figure 2a. We exploited the time dependence of the thermal sintering process for creating different particle shapes: mostly spherical particles with small flat patches (Figure 2a) in one extreme and polygonal particle (Figure 2c) with a rhombic dodecahedron shape in the other. Ultimately, the deformation will continue till the particles reach the Voronoi polygons of their colloidal crystal structure that dictates the symmetries of their deformation. Figures 2a-c exemplifies the tunability of the particle shape for different sintering times: 2-3, 5, and 10 mins, respectively. In the case of 5 mins sintering time, the particles already started to deform and flatten where the particles were physically in contact with neighboring particles. As a result, after the heat treatment each particle had 12 flat patches on its surface. This can be attributed to fact that the each particle has 12 neighbors, of which six are in the same layer and three are in each of the layers above and below (Figure 2b). As can be seen in Figure 2c already 10 mins of sintering time was sufficient for the particles to reach their polygonal Voronoi shape and to transform into rhombic dodecahedra. To quantify the flatness of the patches and the facets in more detail we characterized the particle morphology using Atomic Force Microscopy (AFM) (Figure S1) in tapping mode. These measurements confirm that the patches and the facets are flat and have a root mean-squared roughness of about 9 nm and 4 nm respectively (see supplementary information). Thus, the particle shape only depends on the number and arrangement of neighboring particles that are in direct contact with the particle, and on the heating time. For example, we observed the particles that were in contact with the wall obtained a different shape (see Figure S2). If the particles are ordered in a face-centered-cubic (FCC) array and uniformly deformed for long enough, the resulting particles will be regular rhombic dodecahedra (see inset of Figure2c). Otherwise spheres with a similar symmetric distribution of flat patches will result where the patch diameter can be tuned by time. Similarly, particles with a simple cubic symmetry crystals are predicted to transform into cubes, body centered cubic packed particles will turn into truncated octahedral shapes, and hexagonal close packed particles will transform into hexagonal prism shapes. As a proof of concept, we sintered a body-centered tetragonal (bct) crystal[15,19] which is formed by induced dipolar interactions in the presence of an external electric field (see supplementary information).The resulting particle shape as shown in Figure 2d.

Although the particles completely deformed until the packing fraction nearly reached unity for the crystals shown in Figures 2 c&d, filling all interstitial space in the crystal, it was still possible to resuspend the particles as individual entities by sonicating the annealed crystals for about 10-15 mins at room temperature in CHB. Apparently, the stabilizer forms an effective protective layer. Since the backbone of the stabilizer is a long PMMA chain its covalent bonding to the particle likely involves bonding with several PMMA-molecules in the outer layer of the particle. This would effectively make the stabilizer a cross-linked layer. Additionally, this cross-linked layer completely prevented the inter-diffusion of PMMA polymer chains between the neighboring particles during the thermal annealing and thus the resulting surfaces of the particle remained remarkably smooth as seen in the inset of Figure 2c. Indeed, sterically stabilized but unlocked particles were unable to separate into the individual entities (see Figure S3).

Clearly, for individual particles the equilibrium shape is spherical shape. Fortunately, even after 2-3 weeks of being dispersed in CHB at room temperature the particles still retained their polygonal shape with sharp edges as becomes apparent after drying the particles then examining them with scanning electron microscopy (see Figure S4). We characterized the morphology of the polyhedral particles in terms of the distribution of their edge lengths ($a$), and the percentage of particles with the intended shape present in the sample from SEM images using iTEM software (Olympus Soft Imaging Solutions). The edge length of the particles was measured to be 0.676 ± 0.06 µm with a polydispersity of 8.9 %, and 84% of particles had the rhombic dodecahedron shape. The yield of the rhombic dodecahedron shaped particles is less than 100% mostly because of stacking errors and other defects and can be drastically improved if the level of pure FCC phase is increased for instance by making the colloidal crystals using a template (colloidal epitaxy).[20]

Because of the stability of our deformed particles in CHB these systems can therefore be both index and density matched, so that bulk measurements of self-assembly in real-space are possible. Several recent simulation studies reported on the mesophase behavior of space-filling hard-polyhedrons, namely, truncated octahedrons, rhombic dodecahedrons, hexagonal prisms, cubes, gyrobifastigiums and triangular prisms.[1e,h] For these particle systems the formation of various new liquid-crystalline and 3D rotator or plastic crystalline phases has been predicted at intermediate volume fractions in simulations.[1e,h] Here we show the phase behavior of our sterically stabilized and charged rhombic dodecahedron particles, at different salt concentrations to modify the Debye screening length ($\kappa^{-1}$, measure for the range of repulsion) as compared to the particle size. When the potential between the particles is made soft ($\kappa R \approx 1$, long-ranged screened electro-static repulsions) by dispersing them in a de-ionized solvent (CHB), crystallization occurs at a small volume fraction ($\varphi$). Figure 3a clearly illustrates the softness and the hexagonal positional order of the crystalline particles in the plane (111). Due to the finite resolution (about 250 nm in the imaging plane) of the confocal microscope it is hard to visualize the detailed features of the particle shape. However, it is still clearly visible in Figure 3a that the particles are indeed non-spherical. Due to the long-ranged repulsive interactions between the particles they undergo free 3D rotations on their lattice positions (see supplementary movie1). Therefore, the time-averaged image of this rotator phase over a period of 180s (Figure 3b) shows the polygons as spherical objects. Moreover, the experimental $g(r)$ is in agreement with a 2D hexagonal crystal layer as shown in the Figure 3c. In Figure 3d, the vertical $xz$-slice reveals the *ABC* stacking sequence indeed indicating the FCC symmetry of the crystals. The rotator phase is the phase where the particles are randomly oriented with respect to each other (no long-range orientational order), but the centers of mass of the particles are ordered in a crystalline structure (long-range positional order).[21]

To shed light on how the range of the repulsive double-layer interactions influences the phase behavior, we decreased the electric double layer thickness around the particles by adding a salt like tetrabutylammonium bromide. To achieve hard particle interactions we dispersed the particles in salt saturated (260 µM) CHB. We estimated the Debye screening length ($\kappa R \approx 20$, $R$ is radius of the particle) corresponding to this salt concentration. As a result of hard particle interactions the 3D rotator phase transformed into a defective nonrotator crystalline phase as shown in Figure 3e. If the particles were not given enough time to rearrange themselves then the particles completely lost their long-range positional order and the system became a rotationally disordered glass as shown in Figure 3f. We achieved this state by centrifugation of a dilute ($\varphi \approx 0.08$) sample for 40mins at 2000 rpm. When the particles were dried on a flat substrate we observed orientationally disordered FCC crystals as shown in Figure 3g. It is worthwhile to mention that the particles were not arranged with their flat facets in a side by side fashion because the drying induced forces were much stronger than thermal energy, and that, therefore, the particles did not have time to reorient their facets.

Finally, we show that facets can induce directionality in the inter particle interactions by means of depletion attractions by adding non-adsorbing polymer *(*1.0 wt% of polystyrene polymers, more details see supplementary information*)* to the system. It is clear from the possible increase in overlap volumes that the depletion force should be much more effective between flat interfaces than it is between curved interfaces.[3,12] As a consequence, aggregates composed of short linear segments and closed loops were observed at a low particle concentration (Figure 4a) while the 3D network of gel (Figures 4b&c) was observed at a high particle concentration.

In conclusion, we developed an effective, yet simple method for the fabrication of polyhedron-shaped polymeric particles that reflect the local Voronoi cell of the sphere packings the original polymer spheres were in. Moreover, our method also enables us to tune the particle shape by varying the sintering time. We were able to make round particles with flat patches by stopping at the early stages of the heat induced deformation process. Our method is quite general since it relies only on an initial 3D assembled structure that comprised of spherical particles for which several crystal structures have already been realized. It is a bulk method can easily give yields of grams of particles. As an illustration we demonstrated the phase behavior of rhombic dodecahedron particles at different salt concentrations. We observed the 3D rotator phase in the low salt limit, whereas the particles assembled into a non-rotator phase in the high salt limit. We further demonstrated increased directional interactions by means of depletion attraction between the flat facets of the particles. Although our method was demonstrated for PMMA, but as it only makes use of the surface tension of particles and the fact that exchange of polymers can be drastically reduced by crosslinking it should also apply to other cross-linked polymer systems. As an illustration of this we demonstrated that our procedure indeed also works with cross-linked polystyrene (PS) particles (Figure S5, more details see supplementary info). In addition, we would like to mention that the focus of the present paper is on particles because our intension to use them in real-space confocal studies, but should work just as well with particles with sizes all over the colloidal domain as surface tensions of even particles with a size of 10 nm is still many times $k_BT$. Because of its simplicity, our thermal annealing method opens the possibility for a wide range of amorphous polyhedral particles shapes. We believe that the characteristics of dense packings and the larger interfaces between the facets make the polyhedral particles suitable building blocks for new materials design with respect to their spherical counter-parts. This makes it interesting for both fundamental and applied studies to utilize these new polyhedral particle properties.

## Experimental Section

PMMA particles were synthesized by dispersion polymerization, covalently labelled with the fluorescent dye 7-nitrobenzo-2-oxa-1, 3-diazol (NBD) or rhodamine isothiocyanate (RITC), and sterically stabilized with poly(12-hydroxystearic acid), which was grafted on to a PMMA backbone forming a combgraft steric stabilizer.[16] We used suspensions of 2.6 µm diameter PMMA spheres. The particle size and polydispersity (4 %) were determined by static light scattering (SLS). The particles were dispersed in cyclohexyl bromide (CHB, Fluka), saturated with tetrabutylammonium bromide (TBAB, Sigma). In this dispersion, the particles were nearly refractive index-matched, and they behaved like hard-spheres[19]. We used rectangular capillaries of 0.1 *mm* X 1.0 *mm* and 0.1 *mm* X 2.0 *mm* (VitroCom, UK). After filling the cell with the colloidal suspension, we sealed both ends of the capillary with UV-curing optical adhesive (Norland no.68) and cured the glue with UV-light (λ = 350 nm, UVGL-58 UV lamp). We studied particle dynamics by means of confocal laser scanning microscopy (Leica TCS SP2). We estimated the Debye screening length of our suspensions by measuring the conductivity of the CHB (with a Radiometer analytical CDM 230 conductivity meter) and then applying Walden's rule.[19] We used a sonicator of Branson, Model 8510. We imaged the dried samples with a scanning electron microscopy (FEI Phenom scanning electron microscope). We investigated particle's morphology with atomic force microscopy (Dimension 3100, Bruker).

## Acknowledgements

We thank J. C. P. Stiefelhagen for spherical PMMA particle synthesis and useful discussions; J. D. Meeldijk, and B. Peng for SEM measurements; J. Gerritsen, and S. R. Deshpande (Radboud University, Nijmegen) for AFM measurements, and M. Hermes for fruitful discussions.

**Keywords:** Polyhedral particles • Rotator phases • Self-assembly • Colloids • Patchy particles

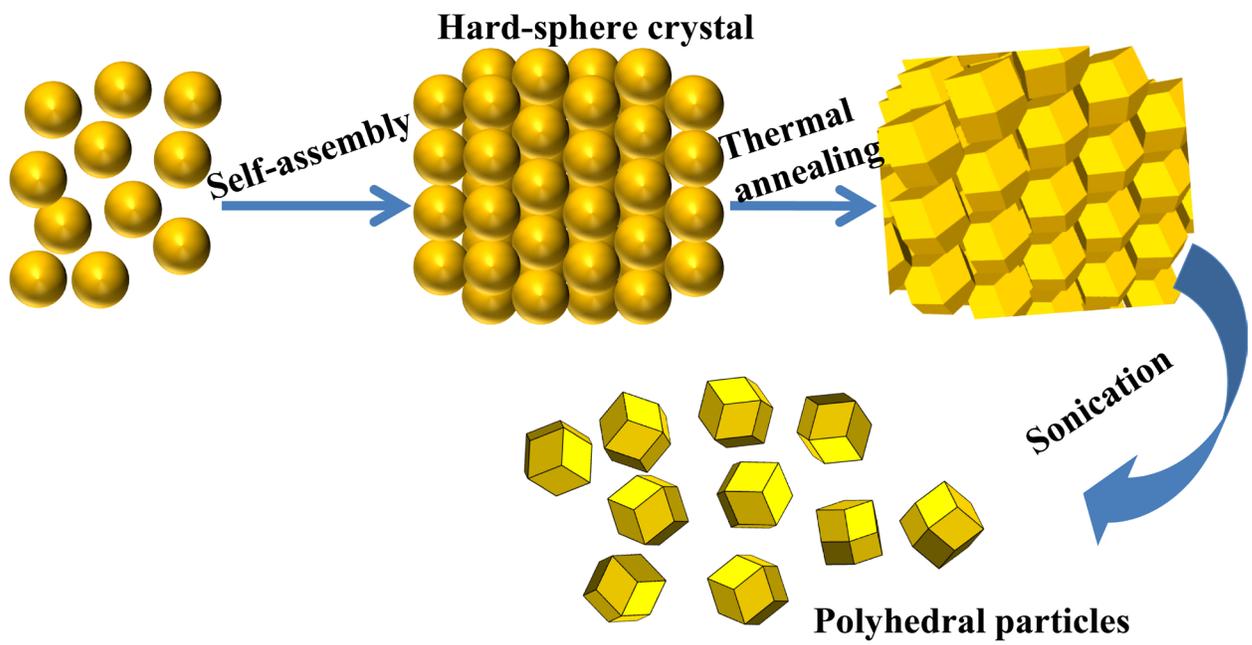

**Figure 1**. Polyhedral particles fabrication. Schematic diagram illustrating the steps involved in the preparation of polyhedral particles.

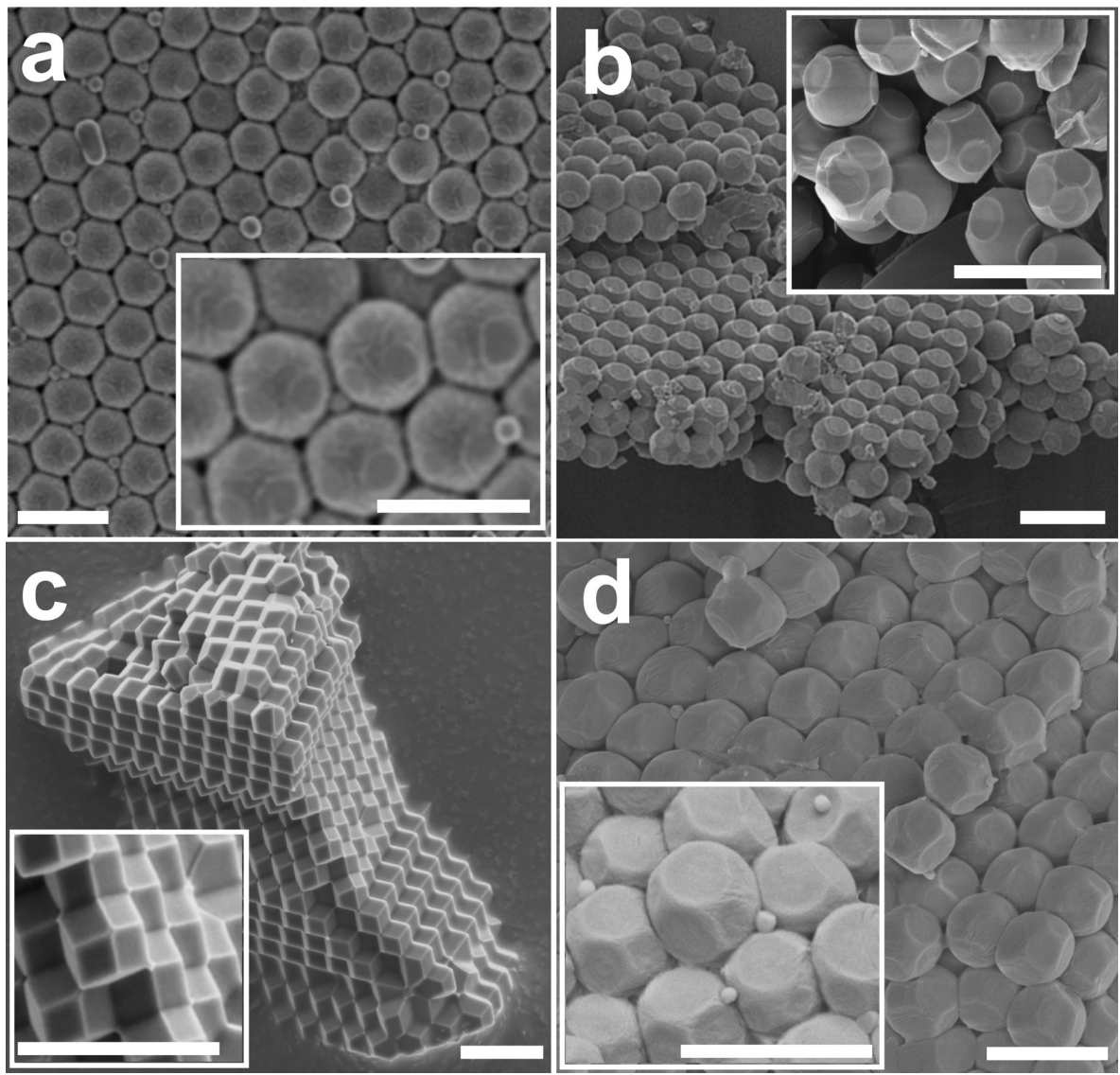

**Figure 2**. Scanning electron micrographs (SEM) of dried polyhedral PMMA particles. **a-c,** Heat treated and the dried hard-sphere FCC crystals of PMMA particles that were annealed at 75 ºC for different heating times ($t_h$) in the presence of the solvent: (a) $t_h$ = 2-3 mins, (b) $t_h$ = 5-6 mins, (insets show magnified views of patchy particles), and (c) $t_h$ = 10-12 mins (inset reveals a magnified view of the rhombic dodecahedron shape). **d,** Heat treated BCT crystal. SEM clearly revealing the characteristic bct stacking: a square arrangement of spheres is obtained perpendicular to an applied electric field (inset shows a magnified view of the particles). Scale bars are 5 µm.

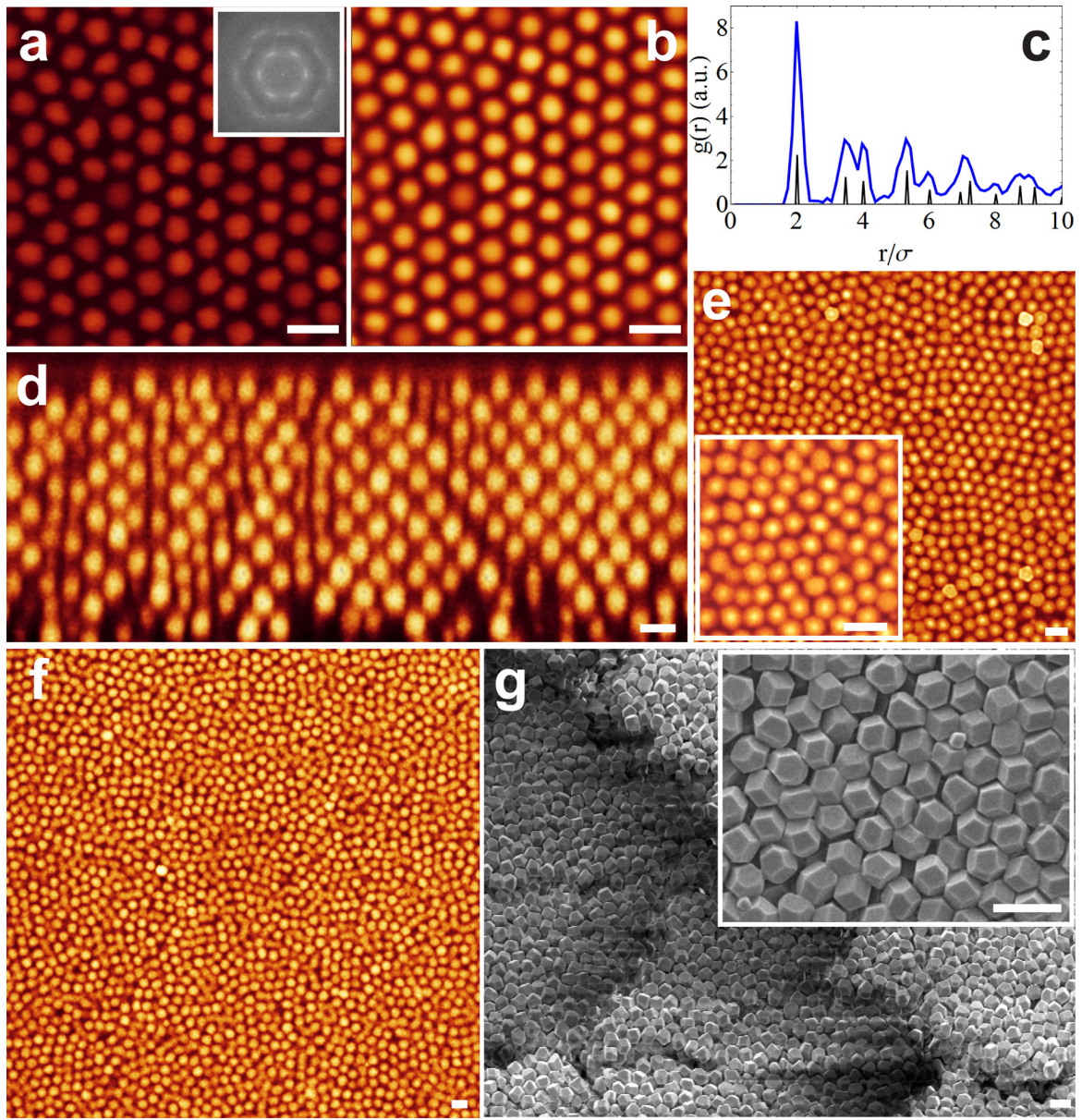

**Figure 3**. Plastic crystals, or rotator phases, of rhombic dodecahedron PMMA particles. **a,** Confocal *xy* snapshot of the FCC (111) plane of a rotator phase of PMMA particles in de-ionized CHB, inset Fourier transform calculated from the real space image (*xy*). **b,** Time-averaged *xy* confocal image over 180 s. **c,** In-plane (2D) radial distribution function (*g(r)*, plotted against *r/σ* where $\sigma \neq 2R$ is mean interparticle distance) calculated from the tracked particle co-ordinates. The experimental *g(r)* was compared with a theoretically calculated 2D hexagonal lattice. **d,** *xz* scan of rotator phase. **e,** Confocal image of the FCC rotator phase of hard-PMMA particles in salt saturated CHB. **f,** Confocal image of rotationally disordered glass. **g,** Scanning electron micrograph of rotationally disordered FCC crystal on a substrate. Scale bars are 5 μm.

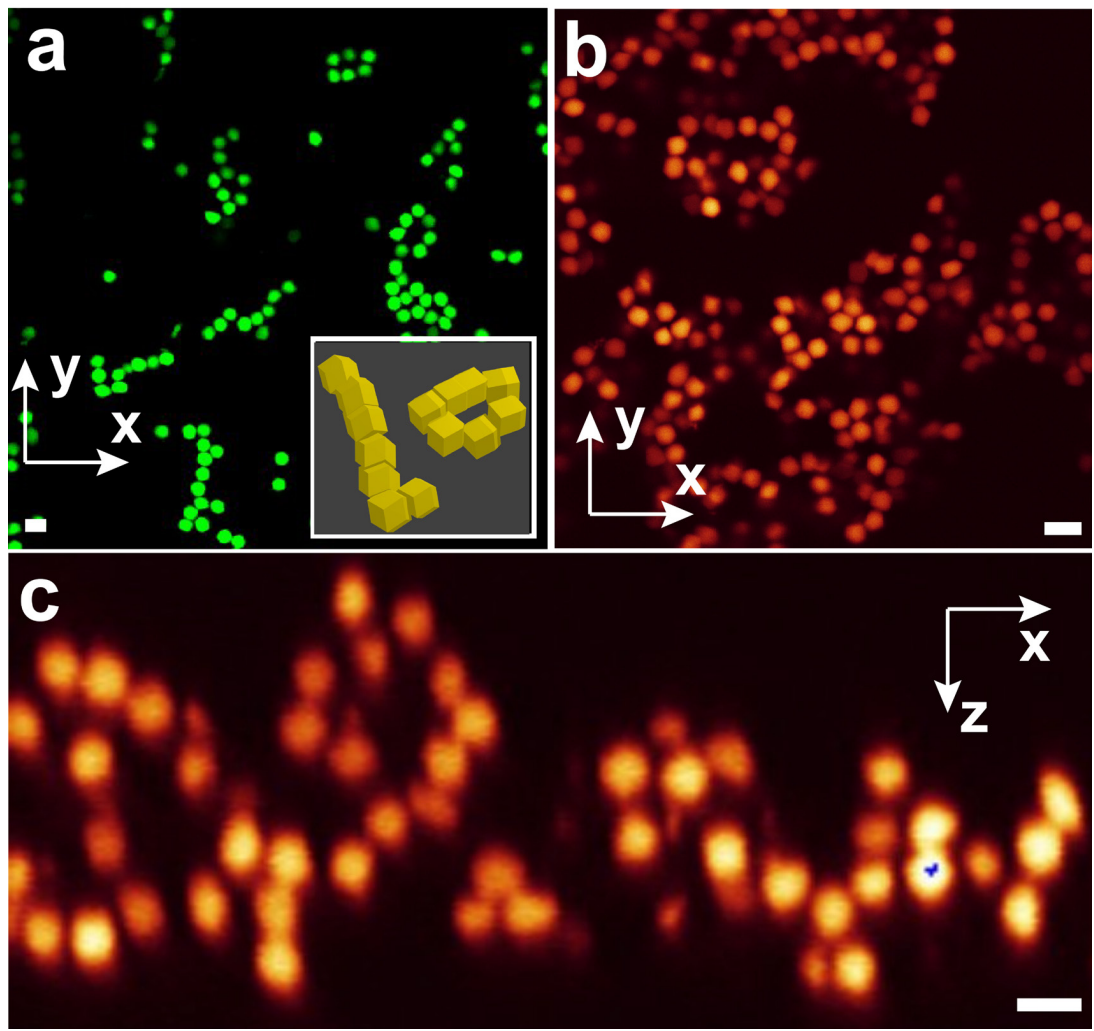

**Figure 4**. Confocal micrographs of clusters of particles, by depletion attractions. **a,** xy confocal micrograph shows particles aggregated in short linear segments and closed loops at low particle concentrations. A conjectural schematic interpretation of internal structures of linear and closed loops is shown as inset. **b&c,** xy and xz confocal images reveal a network of particles at high particle concentrations. Scale bars are 5 μm.